\documentstyle[PASJadd, epsf]{PASJ95}
\draft

\begin{document}
\setcounter{page}{1}

\title{Deep-Imaging Observations
of a Candidate of an Absorbed QSO at $z=0.653$, AX~J131831+3341}

\author{Masayuki {\sc Akiyama},$^{1}$
Kouji {\sc Ohta},$^{2}$\thanks{Visiting astronomer of the University of
Hawaii $88^{\prime\prime}$ telescope.}
Naoyuki {\sc Tamura},$^{2}$\footnotemark[1]
Mamoru {\sc Doi},$^{3,4}$ \\
Masahiko {\sc Kimura},$^{5}$
Yutaka {\sc Komiyama},$^{1}$
Satoshi {\sc Miyazaki},$^{6}$
Fumiaki {\sc Nakata},$^{3}$ \\
Sadanori {\sc Okamura},$^{3,4}$
Maki {\sc Sekiguchi},$^{7}$
Kazuhiro {\sc Shimasaku},$^{3,4}$
Masafumi {\sc Yagi},$^{8}$ \\
Masaru {\sc Hamabe},$^{9}$
Michitoshi {\sc Yoshida},$^{10}$
and
Tadahumi {\sc Takata},$^{1}$
\\ [12pt]
$^{1}$ {\it Subaru Telescope, National Astronomical Observatory of Japan,}\\
{\it 650 North A'ohoku Place, Hilo, HI 96720, U.S.A}\\
{\it E-mail(MA): akiyama@naoj.org} \\
$^{2}$ {\it Department of Astronomy, Faculty of Science,
Kyoto University, Kyoto 606-8502} \\
$^{3}$ {\it Department of Astronomy, School of Science,
The University of Tokyo, Bunkyo-ku,
Tokyo 113-0033} \\
$^{4}$ {\it Research Center for the Early Universe, School of Science,}\\
{\it The University of Tokyo, Bunkyho-ku, Tokyo 113-0033}\\
$^{5}$ {\it Department of Physics, School of Science, The University of Tokyo,
Bunkyo-ku, Tokyo 113-0033}\\
$^{6}$ {\it Advanced Technology Center, National Astronomical Observatory,
Mitaka, Tokyo 181-8588}\\
$^{7}$ {\it Institute for Cosmic Ray Research, The University of Tokyo,
Tanashi, Tokyo 188-8502}\\
$^{8}$ {\it Optical and Infrared Astronomy Division, National Astronomical
Observatory,} \\ {\it Mitaka, Tokyo 181-8588}\\
$^{9}$ {\it Institute of Astronomy, School of Science, The University of Tokyo,
Mitaka, Tokyo 181-0015}\\
$^{10}$ {\it Okayama Astrophysical Observatory, National Astronomical
Observatory,} \\ {\it Kamogata-cho, Okayama 719-0232}\\
}

\abst{
The results of deep-imaging observations of a candidate of an absorbed
QSO at $z=0.653$, AX~J131831+3341, are presented.
AX~J131831+3341 was found during the 
course of optical follow-up observations
of the ASCA Large Sky Survey, and
has an X-ray luminosity of $10^{45}$ erg s$^{-1}$ (2--10~keV), which
corresponds to those of QSOs.
Its optical spectrum shows no significant broad H$\beta$ emission
line, suggesting that the object is an absorbed QSO.
Deep $R$ and $V$ band images reveal the presence of
a point-like nucleus and an asymmetric extended component.
The nuclear component has a blue color, and the optical magnitude 
is much fainter
than that expected from the observed X-ray flux for typical type-1 AGNs.
These photometric properties
and the presence of broad Mg {\scriptsize II} 2800 {\AA}
emission can be explained simultaneously if the observed nuclear light is
dominated by scattered nuclear light, though there is a possibility that
the nuclear component is a slightly absorbed nucleus if its intrinsic
X-ray to optical flux ratio is the largest among X-ray selected AGNs.
The size of the extended component, which is thought to be 
the host galaxy of the QSO, is larger than those of normal disk galaxies
at $z=0-0.75$, and the absolute magnitude 
is similar to those of the brightest host galaxies
of QSOs at redshifts smaller than 0.5.
The $V-R$ and $R-I$ colors of the component are consistent with a
1 Gyr-old stellar population model without absorption.
}

\kword{galaxies: active --- galaxies: individual (AX~J131831+3341)
--- galaxies: photometry --- quasars}

\maketitle
\thispagestyle{headings}

\section{Introduction}

AX~J131831+3341 is an AGN at a redshift of 0.653, discovered during the
optical identification of hard X-ray sources
in the ASCA Large Sky Survey
(LSS; Akiyama et al. 2000, hereafter Paper I).
The optical spectrum of the object
shows strong emission lines, such as
broad Mg {\scriptsize II} 2800 {\AA}, narrow
[O {\scriptsize II}] 3727 {\AA}, and narrow
[O {\scriptsize III}] 5007 {\AA}, but
no broad H$\beta$ emission line.
Its small H$\beta$-to-[O {\scriptsize III}] 5007 {\AA} equivalent
width ratio [log(H$\beta$/[O {\scriptsize III}])$=-0.54$]
is comparable to those of Seyfert 1.8--2 galaxies (Winkler 1992).

The X-ray flux of the object is $5.8 \times 10^{-13} \ {\rm erg} \
{\rm cm}^{-2} \ {\rm s}^{-1}$ in a 2--10~keV band
and the luminosity is estimated to be $10^{45} \ {\rm erg} \ {\rm s}^{-1}$,
which is as large as the luminosity of the knee of the
AGN luminosity function in the 2--10~keV band at $z \sim 0.6$
(e.g., Boyle et al. 1998),
and corresponds to the luminosities of QSOs.
The observed X-ray spectrum of the object in a 0.7--10~keV band
is described by intrinsic absorption with
a hydrogen column density of
$N_{\rm H}=6.0_{-4.2}^{+4.4} \times 10^{21}$ cm$^{-2}$
and an intrinsic photon index of 1.7, which is typical for
broad-line AGNs.
The hydrogen column density corresponds to the lower edge of
the column density distribution of Seyfert 1.8--1.9 galaxies
(Risaliti et al. 1999).

AX~J131831+3341 is detected as a point source
in the FIRST radio source survey
(Becker et al. 1995)
in the 1.4 GHz band with a flux density of 2.0 mJy;
the calculated radio power of the object, 
$L_{\rm 1.4 GHz}$, is  $4 \times 10^{24}$ W  Hz$^{-1}$.
The X-ray to radio-flux ratio of the object is 100-times
larger than those of radio-loud AGNs, and similar to the
upper envelope of radio-quiet AGNs in the ASCA LSS sample (Paper I).
Therefore, AX~J131831+3341 is a good candidate of a radio-quiet
absorbed luminous AGN (absorbed QSO) at an intermediate redshift.

Absorbed QSOs
whose luminous nuclear optical emissions are obscured make it
possible to detect faint and small nebulosities around QSOs and
provide a unique opportunity
to examine the nature of host galaxies of a radio-quiet QSO population.
In fact, an optical image of the object taken without a filter
for object acquisition during a spectroscopic observation shows a
faint nebulosity around a point-like component.
The faint nebulosity extends
$\sim 7^{\prime\prime}$ from the nucleus (Paper I).
Since most of host galaxy imagings of radio-quiet QSOs
by Hubble Space Telescope are limited below a redshift of $\sim 0.5$
(e.g., Bahcall et al. 1997; Boyce et al. 1998; Hooper et al. 1997;
McLure et al. 1999), the X-ray selected absorbed QSO is a
precious site for studying host galaxies of radio-quiet QSOs,
especially in the intermediate-to-high redshift universe.
In this paper, we present results of optical photometric and
deep imaging observations of AX~J131831+3341, and discuss
the possibility that the object is an absorbed QSO as well as the
properties of its host galaxy.
Throughout this paper, we use $q_0=0.5$ and $H_0=50$ km s$^{-1}$ Mpc$^{-1}$.

\section{Observations}

Deep optical imaging observations were made with the Suprime-Cam
attached to the Cassegrain focus of the 8.2 m Subaru telescope on 1999
May 13, June 11, and June 12 
during the first-light phase of the telescope.
The camera consisted of 6 $4096 \times 2048$ CCD chips.
For the observation, a SITe $4096 \times 2048$ chip was used.
The pixel scale of the camera was $0.\!^{\prime\prime}03$ pixel$^{-1}$.
The seeing condition during the observation was not very good
(FWHM is $\sim 0.\!^{\prime\prime}7$).
Three 600 s exposures in the $R$ band and two 900 s exposures
in the $V$ band were obtained.
Unfortunately, because the object fell at an edge of the CCD in
one frame in each band, the northern part of
the image in the $R$ band and the southern part of the image in
the $V$ band were observed with shorter effective exposure times than the
other region.

Photometric calibrations of the $V$ and $R$ band images were performed
based on optical photometric observations made with the Tektronix
$2048 \times 2048$ CCD at the University of Hawaii (UH) $88^{\prime\prime}$
telescope on 1999 March 6.
We obtained Mould $V$ and $R$ band images with exposure times of
450 s and 1500 s, respectively.
During the observing run, 50 Landolt's standard stars
(Landolt 1992) were observed.
The seeing size was $\sim 0.\!^{\prime \prime}8$ and
the pixel scale was $0.\!^{\prime\prime}22$ pixel$^{-1}$.
In the observation, we also obtained Mould $B$ and $I$ band images
with 600 s and 300 s exposures, respectively.

All of the deep imaging and photometric data were
analyzed using IRAF.
Bias subtraction and flat-fielding with dome flats were performed.
Based on the scatters of the countrate-to-magnitude conversion factors
of the observed Landolt's standard stars,
the uncertainties in the photometric calibrations were estimated to be
0.03, 0.06, 0.03, and 0.06 mag in the $B$, $V$, $R$, and $I$ bands,
respectively.
The photometric calibrations of the
deep Subaru images were made with bright stars whose magnitudes
were measured in the UH $88^{\prime\prime}$ images.

\section{Results}
\subsection{Morphology}

The deep $R$ band image of AX~J131831+3341 is shown in figure 1.
We binned the original image in $4 \times 4$ pixels.
The standard deviation in the sky region of the
$V$ and $R$ band deep images are 0.03 and  0.04
counts s$^{-1}$  per $4 \times 4$ binned pixels, which
correspond to 26.47 and 26.07 mag
arcsec$^{-2}$, respectively.
In figure 1,  we show a sky-subtracted image of the object
with the surface brightness ranging
from $-3$ times the standard deviation to
$+9$ times the standard deviation as a gray scale
with a linear scale.

The galaxy has a point-like nucleus and a peculiar extended component.
The profile of the nucleus is well represented by the point spread function (PSF) in the image.
The peculiar structure extends $\sim 8^{\prime\prime}$ away
from the nuclear component.
The structure has an asymmetric feature:
it extends more in the southeastern direction than
in the northwestern direction. In the southeastern region,
there is a gap running from the center to the edge of the component.
An extended structure is also detected in the $V$ and $I$ band
images.


\subsection{Total Magnitudes}

Based on the images
taken with the UH $88^{\prime\prime}$ telescope, the
magnitudes of AX~J131831+3341 are measured
in a $26.\!^{\prime\prime}4$ aperture centered on the nucleus excluding
objects around AX~J131831+3341 and a knot located at $\sim
10^{\prime \prime}$ northwest from the nucleus.
The resulting magnitudes are
21.23$\pm$0.10, 21.08$\pm$0.13, 20.00$\pm$0.10, and 18.96
$\pm$0.13 mag in the 
$B$, $V$, $R$, and $I$ bands, respectively.
The photometric uncertainties include
not only the uncertainties of the flux calibrations (see subsection 2.1),
but also errors of flat-fielding
and uncertainties of sky subtractions
(0.1 mag in the all bands).
In the deep $V$ and $R$-band images obtained with Subaru,
the total magnitudes of AX~J131831+3341 were measured to be
21.06$\pm$0.1 and 20.04$\pm$0.1 mag in the $V$ and $R$ bands with
the same aperture, respectively.
These magnitudes agree with those derived from UH photometric
observations.
The results are summarized in table 1.
Hereafter, we use the $V$ and $R$-band magnitudes from Subaru
observations and the $B$ and $I$-band magnitudes from UH $88^{\prime\prime}$
observations.


\subsection{Colors and Magnitudes of the Extended and the Nuclear Component}

To evaluate the magnitudes and colors of the extended component,
we measured them within regions A and B shown in figure 1
using the deep Subaru $V$ and $R$-band images and the UH
$88^{\prime\prime}$ $I$-band image.
The magnitudes in region A (B) are 22.91 (23.24) mag in the $V$ band,
21.90 (22.14) mag in the $R$ band, and 21.10 (20.92) mag in the $I$ band.
The uncertainty of the photometry is 0.1 mag in the $V$ and
$R$ bands and 0.15 mag in the $I$ band.
Thus, the colors in region A (B) are estimated to be
$V-R$ of 1.01 (1.10) $\pm$ 0.14 mag and
$R-I$ color of 0.80 (1.22) $\pm$ 0.18 mag.

To derive the photometric parameters of the nuclear and extended
components separately, we deconvolved the two components using the
surface-brightness profile of the object.
Since the object has asymmetric and complex features, we decided to
model the surface-brightness distribution of the galaxy to obtain
a rather symmetric and smooth brightness distribution as follows.
We first fitted ellipses, of which the centers, position angles,
ellipticities, and surface brightnesses were free parameters, to
the isophotes of the object, using the {\bf ellipse} command of the
{\bf isophot} package in the IRAF.
The sampling step of the ellipses changes with the semi-major axis
length; the step size is taken to be one twelfth of the semi-major
length of each ellipse.
Next, based on the fitted parameters, we constructed a model image
of the galaxy with {\bf bmodel} command of the above package.
The $R$-band image of the constructed model and the residual image
after subtracting the model from the original image are shown
in figure 2.
(Figure 2 is shown in the same count rate range as that used in figure 1.)
As can be seen in figure 2, 
the constructed model well describes the overall
shape of the original image.
The total magnitude of the residual image is 22.0 mag, which corresponds to 16\% of the total
magnitudes.
The method was applied not only to the deep $V$ and $R$ images, but also
to the $I$-band image in which the extended component is also detected.

The surface brightness profiles are derived as sections of
the model image along the major-axis (position angle of 114$^{\circ}$),
and are shown in figure 3.
The derived profiles are consistent with the cross sections along
the major axis of the object, though they are not as smooth as
expected from the image.
Figure 3 shows that the extended component follows the exponential
profile, which is  typical for galactic disks.
In both the $V$ and $R$-band profiles, in the inner region
($r \leq 4^{\prime\prime}$), the profiles in the southeastern and
northwestern directions agree with each other within the uncertainties of
the model fittings, while in the outer region ($r \geq 4^{\prime\prime}$)
the southeastern component is brighter than the northwestern component.
The exponential law is fitted to the profiles in the regions from
$1.\!^{\prime\prime}5$ to $4^{\prime\prime}$ in both the $V$ and $R$-bands.
The best-fit profiles are
$\mu_V  = 24.73 + 0.24 \ r(^{\prime\prime})$  and
$\mu_R  = 23.41 + 0.36 \ r(^{\prime\prime})$ (mag arcsec$^{-2}$) for
the southeastern direction and
$\mu_V  = 24.52 + 0.34 \ r(^{\prime\prime})$ and
$\mu_R  = 23.61 + 0.24 \ r(^{\prime\prime})$ (mag arcsec$^{-2}$) for
the northwestern direction.
The fitting uncertainties of the central surface brightness and
the disk scale lengths are $\sim$ 0.2 mag arcsec$^{-2}$
and $0.\!^{\prime\prime}5 - 1^{\prime\prime}$, respectively.
By subtracting the fitted exponential profiles in the southeastern
(northwestern) direction from the original profile, the profiles of
the nuclear component are obtained in each direction.
Since the obtained profiles agree well with the PSF in each band,
which is derived by averaging the profiles of stellar objects,
we fitted the PSF to the residual profiles.
The dashed lines in figure~3 represent the summed profiles of the 
best-fit PSFs and the best-fit exponential laws in the $V$ and $R$-bands.
Since these deconvolutions reproduce the observed profiles
satisfactorily in each band, we did not undertake an iteration of the
profile-fitting process, and adopted these fitting results as the
final ones.
We applied the same deconvolution method to the $I$-band image,
assuming that its scale length to be the same
as that in the $R$-band image, and that the extended component dominates
the surface brightness in the range between $1.\!^{\prime\prime}5$ and
$2^{\prime\prime}$.

The magnitudes of the resulting nuclear components were calculated to be
22.61 (22.64), 22.26 (22.22), and 21.77 (21.66) mag
in the $V$, $R$, and $I$ bands, respectively.
(The values in parentheses denote those determined from the
northwestern profile.)
Because the central surface brightnesses are dominated by the nuclear
component, the magnitudes derived from the southeastern and
northwestern profiles agree well.
The uncertainty of the magnitude of the nuclear component is dominated by
that of the photometric calibration and sky subtraction
(0.1 mag in the $V$ and $R$ band and 0.13 mag in the $I$ band).
We list the derived values in the southeastern region in table 1.
Hereafter, we refer to these magnitudes as those of
the nuclear component.
Subtraction of the magnitudes of the nuclear component
from the total magnitudes of AX~J131831+3341 gives the magnitudes
of the extended components:
21.36, 20.19, and 19.04 mag in the $V$, $R$, and $I$ bands,
respectively.
The nuclear component contributes $24\%$ in the $V$ band, 
$13\%$ in the $R$ band, and
$8\%$ in the $I$ band of the total flux; thus,
the contribution from the extended component dominates
the total magnitudes of AX~J131831+3341 in these bands.
From these magnitudes, the colors of the extended component are
estimated to be $V-R = 1.17 \pm 0.14$ mag and $R-I = 1.15 \pm 0.16$ mag.


\section{Discussions}
\subsection{Nature of the Nuclear Component}

The nuclear component should be related to the AGN;
strong broad Mg {\scriptsize II} 2800 {\AA} emission line
is detected and the rest-frame equivalent width of the broad line (99$\pm$25 {\AA})
is larger than those of normal QSOs
(50$\pm$29 {\AA} ; Francis et al. 1991),
implying that the nuclear component is dominated by
AGN light, at least in the $B$ band.
Possible explanations for the origin of nuclear light
are that we see either an absorbed AGN directly
or scattered AGN light.

Regarding the first case,
if we assume that the nucleus has the same intrinsic
X-ray to optical flux ratio of soft X-ray selected AGNs/QSOs
[$\log (f_{\rm X_{\rm 0.3-3.5keV}}/f_{V}$) values of
Einstein Medium Sensitivity Survey (hereafter EMSS) AGNs
range from $-1.0$ to $+1.4$ and peak at $+0.2$;
Stocke et al. 1991]
and a typical photon index of type~1 AGNs (1.7),
it should appear with a magnitude of 14.5 -- 20.5 mag in the $R$ band.
If the object has a typical X-ray to optical flux ratio [$\log (f_{\rm
X_{\rm 0.3-3.5keV}}/f_{V})$ = $+0.2$], it
should have a magnitude of $17.5$ mag in the $R$ band.
The observed $R$-band magnitude
of the nuclear component is much fainter than the magnitude range.
It should be noted that
BL Lac objects have a larger $\log (f_{\rm X_{\rm 0.3-3.5keV}}/f_{V})$
value (+0.3 to +1.7; Stocke et al. 1991) than normal AGNs/QSOs;
but, AX~J131831+3341 is not such  an object, because it is not a
radio-loud AGN.
If the absorption causes the faintness of the nuclear component,
an optical extinction larger
than $1.76$ mag at 4000 {\AA} to the nucleus is required.
The optical extinction corresponds to $A_V$ of larger than 1.2 mag,
based on the extinction curve of the Galaxy (e.g., Scheffler,
Els\"asser 1988).
If the object intrinsically has the typical X-ray to optical flux ratio,
the optical extinction to the nucleus is estimated to be $A_V = 3.3$ mag.
The derived range of the optical extinction is consistent with that derived from
X-ray spectral fitting ($A_V = 3.4_{-2.3}^{+2.5}$ mag, if we assume
the correlation of X-ray absorption and
optical absorption in the galactic interstellar gas (e.g.,
Predehl, Schmitt 1995)).
The expected range of colors of the average QSO SED absorbed with
$A_{V}$ between 1.1 mag and 5.9 mag is indicated by the dashed line
 in figure 4.
The $V-R$ and $R-I$ colors of the nuclear component are
marginally consistent with those of an averaged QSO SED
(Francis et al. 1991) if $A_V$ is about 1.0 mag,
but much bluer than the QSO SED with $A_V=3.3$ mag.
Thus, only if the intrinsic X-ray to optical flux ratio of the object
is the largest among the X-ray selected AGNs (and thus in the minimum
extinction case),
the large X-ray to optical flux ratio and blue colors of the nuclear
component can be explained simultaneously.

In narrow-line AGNs, some fraction of the nuclear light is scattered
into our line of sight by dust particles or electrons.
In narrow-line radio galaxies, the fraction is estimated to be
a few percent based on spectropolarimetric observations
(Alighieri et al. 1994).
Such scattered light from an obscured AGN can dominate the optical
light of an object, especially in the wavelength range below the 
rest-frame 4000 {\AA}, if the obscured AGN is intrinsically as luminous
as the QSOs and the host has a red color, like that of 
an elliptical galaxy.
For AX~J131831+3341, assuming that the nucleus has the
same intrinsic optical to X-ray flux ratio of AGNs in the
EMSS sample and that 2 \% of the nuclear light is scattered isotropically,
the scattered component should appear with 18.7 -- 24.7 mag in the
$R$ band.
The observed $R$-band magnitude of
the nuclear component falls in the magnitude range.
The $V-R$, and $R-I$ colors are consistent with
$f_{\nu} \sim \nu^{-1.0}$ and the
$f_{\nu} \sim$ constant model with an $A_V$ smaller than $1$ mag.
Such colors are similar to those of the scattered-light component in
3C radio galaxies (Alighieri et al. 1994).
Therefore, the scattered AGN light can be the origin of the
nuclear component.

The X-ray to Mg {\scriptsize II} 2800 {\AA} flux ratio of
AX~J131831+3341 does not conflict with both models;
the equivalent width of the broad line is similar to
that of normal QSOs and the X-ray to optical flux ratio is
explained by the two models.
A broad Mg {\scriptsize II} 2800 {\AA} emission line
found in narrow-line radio galaxies
(Alighieri et al. 1994; Barcons et al. 1998) and
narrow-line ultra-luminous infrared galaxies (Hines, Wills 1993;
Hines et al. 1995) is thought to originate from scattered nuclear light.
Non-detection of a broad H$\beta$ line can be explained by
dilution of the nuclear emission
by the host-galaxy component in a red wavelength range
(see e.g., figure 5 in Alighieri et al. 1994, figure 3 in
Willott et al. 2000) in the case of the scattered-light origin.

\subsection{Nature of the Extended Component}

The extended component is considered to be the host galaxy of the QSO.
The asymmetric and extended structure of the component suggests
a galaxy interaction or merging in the host galaxy.
The gap, which runs from the center to the edge in the eastern extension,
could be a dust lane in the component, or the component could
consist of two tidal tails.
The morphology of the component is similar to that of an interacting
galaxy, NGC~2992, though the extension of AX~J131831+3341
($\sim 8^{\prime\prime}$ corresponds to 62 kpc at $z=0.653$)
is much larger than that of NGC~2992 (20 kpc away from the
nucleus; Ward et al. 1980).
Similar extended tails can be seen in some Ultra Luminous Infrared
Galaxies (ULIGs, Sanders et al. 1988).
It is also possible that we see an edge-on disk galaxy with an
asymmetric dust lane.


We compared the colors of the extended component
(region A, region B, and total$-$nucleus)
with spectral evolution models of the galaxies in figure 4.
We used two models (Kodama, Arimoto 1997):
1) Elliptical model (plotted with a solid line)
in which star formation occurs during the first 0.353 Gyr with
an initial mass function with a slope of 1.20;
after that the galaxy evolves passively.
The model parameters well reproduce the reddest and
brightest ($M_V=-23$) class elliptical galaxy in the Coma cluster
(Kodama et al. 1998).
2)
Disk model (plotted with a dotted line) is used
in which star formation occurs constantly
with the same initial-mass function as that in the elliptical model.
The colors of these models at ages from 0.01 Gyr to 12 Gyr at a
redshift of 0.653 are shown as tracks.
We mark the positions of models with 6 Gyr age
with tick marks on the tracks of both models. The age
corresponds to that of the universe at a
redshift of 0.653 under the adopted cosmological parameters of
$q_0=0.5$ and $H_0=50$ km s$^{-1}$ Mpc$^{-1}$.
The colors of the extended component fall on the track
of an elliptical model with an age of 0.5 -- 1 Gyr, and
the host galaxy is significantly bluer than the old 6 Gyr
elliptical galaxy model in the $R-I$ color.
Because the 0.5 -- 1 Gyr elliptical model spectrum
resembles that of a post-starburst galaxy, 
AX~J131831+3341 may have a post-starburst galaxy as the host
like ``post-starburst quasar'', UN~J1025$-$0040 (Brotherton et al. 1999),
which is found at a redshift of 0.6344.
If we introduce optical extinction in the host galaxy,
the colors are also explained by
a disk model or
an elliptical or disk model at ages of less than 0.1 Gyr
with an absorption of $A_V = 1 - 3$ mag.

From the $R$-band magnitude, which nearly corresponds to the 
$B$ band in the object rest frame,
the absolute magnitude of the extended component
is estimated to be
$M_B = -22.8$ mag, $M_V = -24.1$ mag, and
$M_R = -23.7$ mag
with
a {\it K}-correction of 0.127 mag in the $R$ band,
$B-R$ color of 0.88 mag, and
$V-R$ color of 0.41 mag in the object rest frame,
based on the 1 Gyr elliptical model.
It should be noted that
since the $B$, $V$, $R$, and $I$ bands in the observer's frame cover
the $B$ and $V$ bands in the object frame, the rest frame
$V-R$ color
is an extrapolation of the best-fit model in the observed range.
The absolute magnitudes are $\sim$ 2 mag brighter
than that of the knee of the
galaxy luminosity function at the object's redshift ($M_B^{*} \sim -21$ mag;
Lilly et al. 1995),
and correspond to the brightest host
galaxies of QSOs at a redshift less than 0.5 observed with the Hubble
Space Telescope ($M_R = -22.5 - -24.5$ mag,
McLure et al. 1999; $M_R = -22.24 - -23.97$ mag,
Hooper et al. 1997; $M_V = -20.5 - -23.5$ mag,
Bahcall et al. 1997).
The absolute magnitude also matches the brightest ULIG in the
intermediate redshift ($z \sim 0.2$) universe ($M_R = -21.9 - -23.7$,
Zheng et al. 1999, assuming $R-I$ color of an elliptical
galaxy as 0.7).
The post-starburst nature could make the host galaxy brighter than
the absolute magnitude of the knee of the galaxy luminosity function
at the object redshift; the 
total magnitude in the observer's frame $R$ band of the 1 Gyr elliptical
 galaxy model is 2.7 mag brighter than that of the 6 Gyr elliptical
 galaxy model having the same mass.

We compare the scale length and the central surface brightness of the
host galaxy with those of normal galaxies,
though the extended component may not be a normal disk of a galaxy.
The disk scale lengths derived from the fitting made in section
3.3 are  $4.\!^{\prime\prime}1$ and $3.\!^{\prime\prime}0$ in $V$ band
and $2.\!^{\prime\prime}8$ and $4.\!^{\prime\prime}2$ in $R$ band
for the southeastern and the northwestern directions, respectively.
These values in the $R$ band correspond to the scale lengths of 22$
\pm 8$ kpc and 33$ \pm 8$ kpc at the object redshift in the
southeastern and northwestern direction, respectively.
The lengths are larger than the largest disk scale length
of nearby galaxies ($\sim $20 kpc; Kent 1985).
The large extended feature resembles that of early-type galaxies
(NGC~3872, NGC~5533) and interacting galaxies (NGC~2770, NGC~5905)
in the sample by Kent (1985).
Based on the observed central surface brightness of the extended component
in the $R$ band, that in the rest-frame $B$ band is estimated to be 21.44
 mag arcsec$^{-2}$ by correcting for the $(1+z)^{-4}$ surface brightness
 dimming ($2.18$ mag arcsec$^{-2}$),
{\it K}-corrections (0.67 mag) and the $B-R$ colors (0.88 mag)
of the 1 Gyr elliptical model.
The brightness corresponds to a typical value of the normal disk
galaxies ($\mu_{\it B}=20 - 23$ mag arcsec$^{-2}$; Freeman 1970;  Kent 1985)
but brightest end of disks with a scale length of $\sim 20$ kpc.
The obtained scale lengths are almost the largest among those of
normal disk galaxies at $z=0.5-0.75$; also,
the central surface brightnesses
are as bright as, or slightly fainter than, those of normal disk
galaxies at $z=0.5 - 0.75$ (Lilly et al. 1998).

In figure 5, we compared the surface brightness profile of
AX~J 131831+3341 with the best-fit profiles of the QSO
host galaxies at $z \sim 0.2$
detected in observations with Hubble Space Telescope
(McLure et al. 1999).
In contrast to the exponential profile of AX~J131831+3341,
most of the observed profiles are well described by the
de Vaucouleurs $r^{1/4}$ law, which is a typical profile
of an elliptical galaxy and a bulge of a spiral galaxy.
The best-fit
profiles were converted from those at a redshift of $\sim 0.2$ to that
at a redshift of 0.653 with surface-brightness dimming of
$(1+z)^{-4}$ and a band-shift effect by using SED of an elliptical
galaxy at $z=0$ ($\sim$ 1 mag in $R$ band),
which represents their $R-K$ colors (McLure et al. 1999).
We convolved these profiles with the PSF in our data.
The extended component of AX~J131831+3341 has a brighter
surface brightness and a larger scale length than
that of the QSO host galaxies.


Many absorbed radio-quiet QSOs, like AX~J131831+3341,
are expected to be found
in optical follow-ups of hard X-ray surveys
with new-generation X-ray satellites, Chandra and XMM-Newton 
observatories.
Information on the morphological, photometric, and spectroscopic
characteristics of the 
host galaxies of absorbed radio-quiet QSOs, especially
at an intermediate-to-high redshift universe, will be an important key
to understand the QSO populations, host galaxies of QSOs and
connection between galaxy evolution and AGN evolution
(e.g., Boyle, Terlevich 1998).

\par
\vspace{1pc}\par
KO and NT appreciate the support from members of the UH observatory
during the imaging observations.
MA acknowledges support from 
Research Fellowships of the Japan Society for the Promotion of Science
for Young Scientists.
The optical follow-up program is supported by grants-in-aid from the
Ministry of Education, Science, Sports and Culture
(06640351, 08740171, 09740173) and from the Sumitomo Foundation.

\clearpage
\section*{References}
\re
Akiyama M., Ohta K., Yamada T., Kashikawa N., Yagi M.,
Kawasaki W., Sakano M., Tsuru T. et al.\ 2000, ApJ 532, 700 (Paper I)
\re
Alighieri S.S., Cimatti A., Fosbury R.A.E.\ 1994, ApJ 431, 123
\re
Bahcall J.N., Kirhakos S., Saxe D.H., Schneider D.P.\ 1997, ApJ 479, 642
\re
Barcons X., Carballo R., Ceballos M.T., Warwick R.S., Gonzalez-Serrano J.I.\
1998, MNRAS 301, L25
\re
Becker R.H., White R.L., Helfand D.J.\ 1995, ApJ 450, 559
\re
Boyce P.J., Disney M.J., Blades J.C., Boksenberg A., Crane P.,
Deharveng J.M., Macchetto F.D., Mackay C.D., Sparks W.B.\
1998, MNRAS 298, 121
\re
Boyle B.J., Georgantopoulos I., Blair A.J., Stewart G.C., Griffiths
R.E., Shanks T., Gunn K.F., Almaini O.\ 1998, MNRAS 296, 1
\re
Boyle B.J., Terlevich R.J.\ 1998, MNRAS 293, L49
\re
Brotherton M.S., van Breugel W., Stanford S.A., Smith R.J.,
Boyle B.J., Miller L., Shanks T., Croom S.M., Filippenko A.V.\
1999, ApJ 520, L87
\re
Francis P.J., Hewett P.C., Foltz C.B., Chaffee F.H.,
Weymann R.J., Morris S.L.\ 1991, ApJ 373, 465
\re
Freeman K.C.\ 1970, ApJ 160, 811
\re
Hines D.C., Schmidt G.D., Smith P.S., Cutri R.M., Low F.J.\
1995, ApJ 450, L1
\re
Hines D.C., Wills B.J.\ 1993, Rev.Mexicana Astron.Astrofis. 27, 149
\re
Hooper E.J., Impey C.D., Foltz C.B.\ 1997, ApJ 480, L95
\re
Kent S.M.\ 1985, ApJS 59, 115
\re
Kodama T., Arimoto N.\ 1997, A\&A 320, 41
\re
Kodama T., Arimoto N., Barger A.J., Arag\'on-Salamanca A.\
1998, A\&A 334, 99
\re
Landolt A.U.\ 1992, AJ 104, 340
\re
Lilly S.J., Schade D., Ellis R., Le Fevre O., Brinchmann J.,
Tresse L., Abraham R., Hammer F. et al.\ 1998, ApJ 500, 75
\re
Lilly S.J., Tresse L., Hammer F., Crampton D., Le F\'evre O.\ 1995,
ApJ 455, 108
\re
McLure R.J., Kukula M.J., Dunlop J.S., Baum S.A., O'Dea C.P.,
Hughes D.H.\ 1999, MNRAS 308, 377
\re
Predehl P., Schmitt J.H.M.M.\ 1995, A\&A 293, 889
\re
Risaliti G., Maiolino R., Salvati M. 1999, ApJ 522, 157
\re
Sanders D.B., Soifer B.T., Elias J.H., Madore B.F.,
Matthews K., Neugebauer G., Scoville N.Z.\
1988, ApJ 325, 74
\re
Scheffler H., Els\"asser H.\ 1988, Physics of the Galaxy and Interstellar
Matter (Springer Verlag, Berlin)
\re
Stocke J.T., Morris S.L., Gioia I.M., Maccacaro T.,
Schild R., Wolter A., Fleming T.A., Henry J.P.\
1991, ApJS 76, 813
\re
Ward M., Penston M.V., Blades J.C., Turtle A.J.\  1980, MNRAS 193, 563
\re
Willott C.J., Rawlings S., Blundell K.M.,
Lacy M.\ 2000, MNRAS in press
\re
Winkler H.\ 1992, MNRAS 257, 677
\re
Zheng Z., Wu H., Mao S., Xia X.-Y., Deng Z.-G.,
Zou Z.-L.\ 1999, A\&A 349, 735
\begin{table*}[t]
\begin{center}
Table~1.\hspace{4pt}Summary of photometry of AX~J131831+3341.\\
\end{center}
\vspace{6pt}
\begin{tabular*}{\textwidth}{@{\hspace{\tabcolsep}
\extracolsep{\fill}}lccccccc}
\hline\hline\\[-6pt]
                      &  $B$   & $V$    & $R$   &  $I$  &
                          $B-V$     &  $V-R$      & $R-I$ \\
                      &  (mag) & (mag)  & (mag) & (mag) &
                            (mag)  & (mag) & (mag) \\[4pt]\hline\\[-6pt]
 Total (from UH$88^{\prime\prime}$)
                      & 21.23 & 21.08 & 20.00 & 18.96 &
                                 0.15 & 1.08 & 1.04  \\
 Total (from Subaru)  & $\cdots$ & 21.06 & 20.04 & $\cdots$ &
                             $\cdots$ & 1.02 & $\cdots$  \\
 Nucleus              & $\cdots$ & 22.61 & 22.26 & 21.77 &
                             $\cdots$ & 0.35 & 0.49   \\
 Total $-$Nucleus     & $\cdots$ & 21.36 & 20.19 & 19.04 &
                             $\cdots$ & 1.17 & 1.15   \\
 Region A             & $\cdots$ & 22.91 & 21.90 & 21.10 &
                             $\cdots$ & 1.01 & 0.80   \\
 Region B             & $\cdots$ & 23.24 & 22.14 & 20.92 &
                             $\cdots$ & 1.10 & 1.22   \\
\hline
\end{tabular*}
\end{table*}

\clearpage
\centerline{Figure Captions}
\bigskip
\begin{fv}{1}
{7cm}
{Deep $R$-band $4 \times 4$ binned image of AX~J131831+3341 with a total
 exposure time of 1800 s.
The field of view of the image is $24^{\prime\prime} \times
24^{\prime\prime}$. North is up and east is to the left.
We show the sky-subtracted image of the object
with a surface brightness ranging
from $-3$ times the standard deviation to
$+9$ times the standard deviation as a gray scale
with a linear scale.
Regions A and B, where the colors of the extended component
are measured, are indicated by rectangles.
}
\end{fv}

\begin{fv}{2}
{7cm}
{(a) $R$-band model image of AX~J131831+3341 constructed from an ellipse fitting
(see text for details).
(b) Residual image after subtracting the model image from the original image.
The field of view of the images and the gray scale levels
are the same as in figure 1.
}
\end{fv}

\begin{fv}{3}
{7cm}
{$V$, $R$, and $I$ band profiles of AX~J131831+3341
along the major axis of the extended component.
The profiles in the 
southeastern and northwestern directions are shown by the
thick and thin solid lines, respectively.
The thick and thin dashed lines for 
the $R$ and $V$ bands represent the sum of
the fitted exponential-law profile and the PSF profile for the
southeastern direction and for the northwestern direction,
respectively.}
\end{fv}

\begin{fv}{4}
{7cm}
{$V-R$ and $R-I$ colors of the nuclear (filled square)
and the extended (region A: filled triangle, region B: filled pentagon,
total$-$nucleus: filled circle)
components of
AX~J131831+3341.
The tracks of the elliptical and disk models
with ages from 0.01 Gyr to 12 Gyr are indicated by the 
thick solid and dotted lines, respectively.
The positions at an age of 6 Gyr are 
indicated by the tick marks on each track.
The open circle represents the 
color of an average QSO SED (Francis et al. 1991).
The pentagons show the colors of a power-law model with
indices ($f_{\nu}=\nu^{\alpha}$)
of $\alpha = -1.0$, $0.0$, $1.0$ from top to bottom.
The dashed line indicates the 
expected colors for the average QSO SED with the
absorption derived from the X-ray spectrum ($A_V = 1.1 - 5.9$ mag).
The thick region of the dashed line shows an absorption larger than
$A_V = 3.3$ mag, which was obtained for the nuclear component by assuming
the typical optical to X-ray flux ratio of X-ray selected AGNs.
The arrow represents the effect of reddening with $A_V$ of 1 mag.}
\end{fv}

\begin{fv}{5}
{7cm}
{$R$-band profiles of AX~J131831+3341
in the southeastern and the northwestern directions
 (thick solid line) compared with those
 of QSO host galaxies at $z < 0.3$ (McLure et al. 1999) (thin solid lines).
The best-fit model profiles of the QSO host galaxies at $z <0.3$ are
convolved with our PSF, and are 
corrected for {\it K}-correction and surface brightness
dimming to compare the profiles at $z=0.653$.}
\end{fv}

\clearpage

 \begin{figure}
 \hspace{2cm}
 \epsfbox{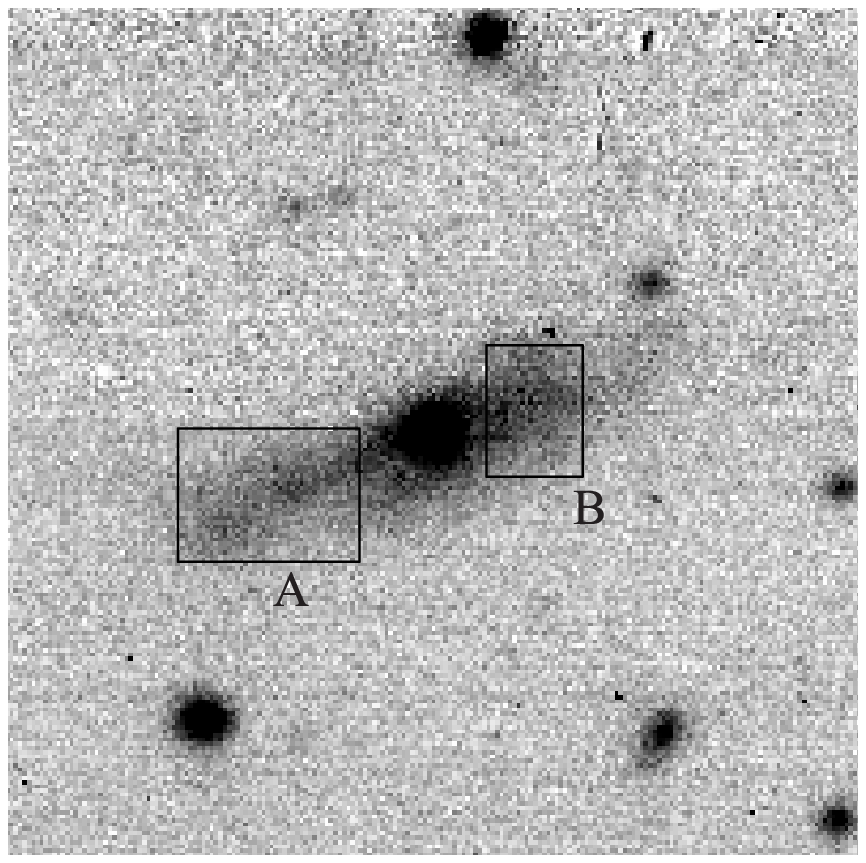} \\
 \caption{}
 \end{figure}

 \begin{figure}
 \hspace{2cm}
 \epsfbox{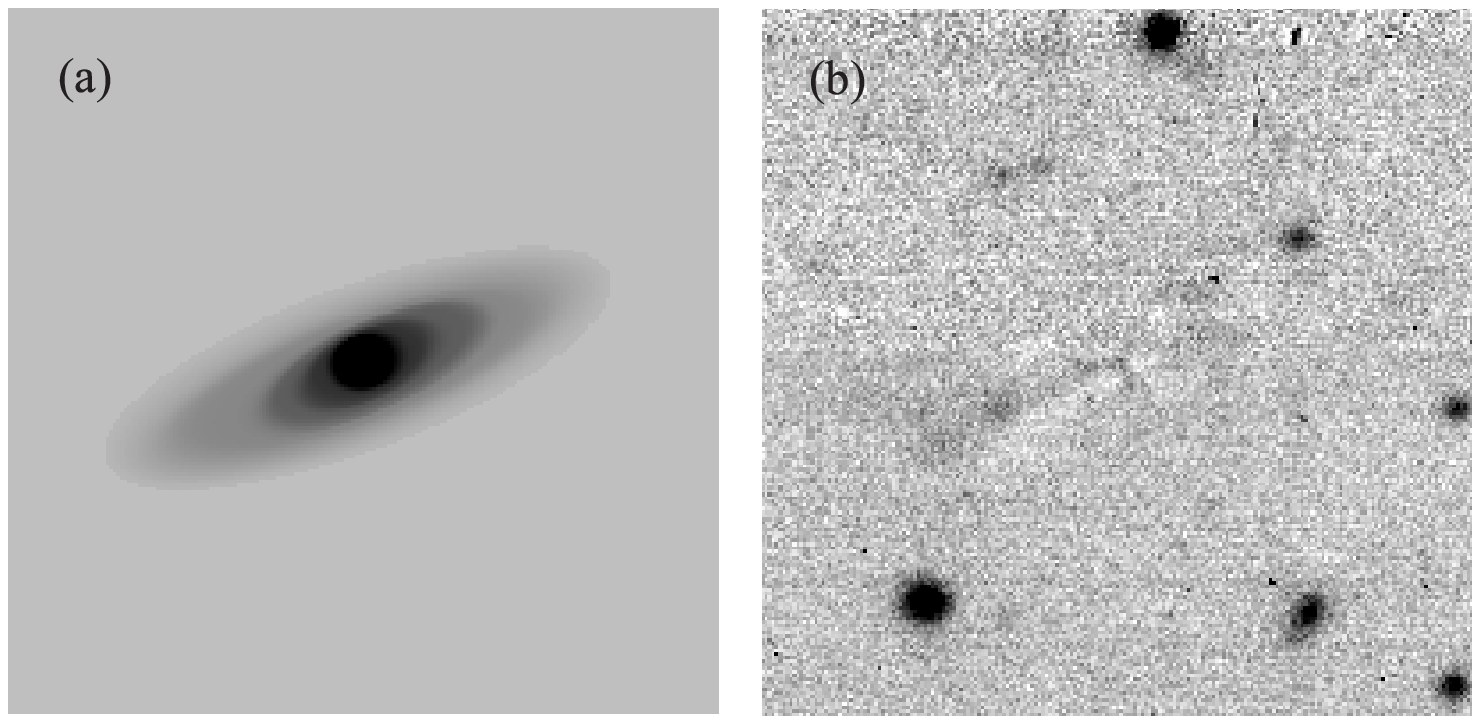} \\
 \caption{}
 \end{figure}
 
 \clearpage
 
 \begin{figure}
 \hspace{2cm}
 \epsfbox{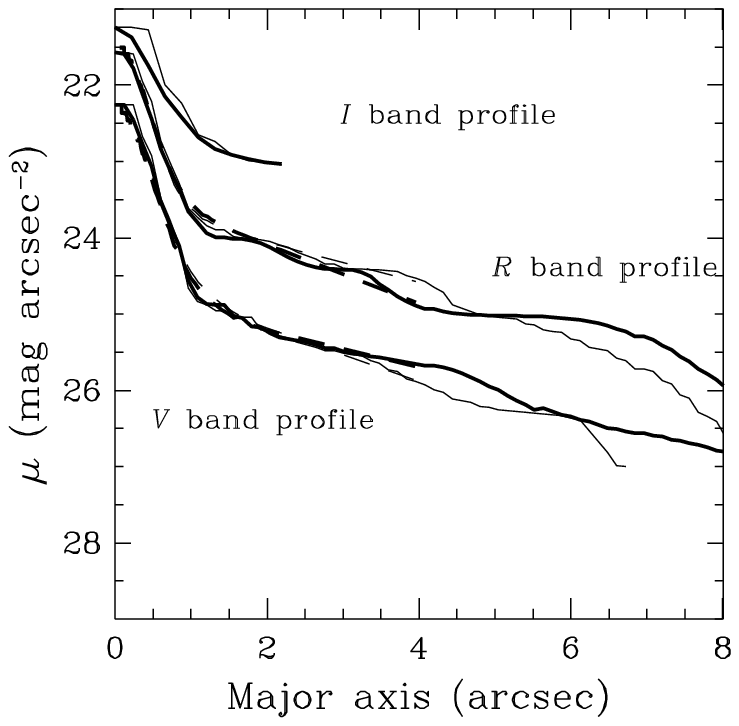} \\
 \caption{
 }
 \end{figure}
 
 \begin{figure}
 \hspace{2cm}
 \epsfbox{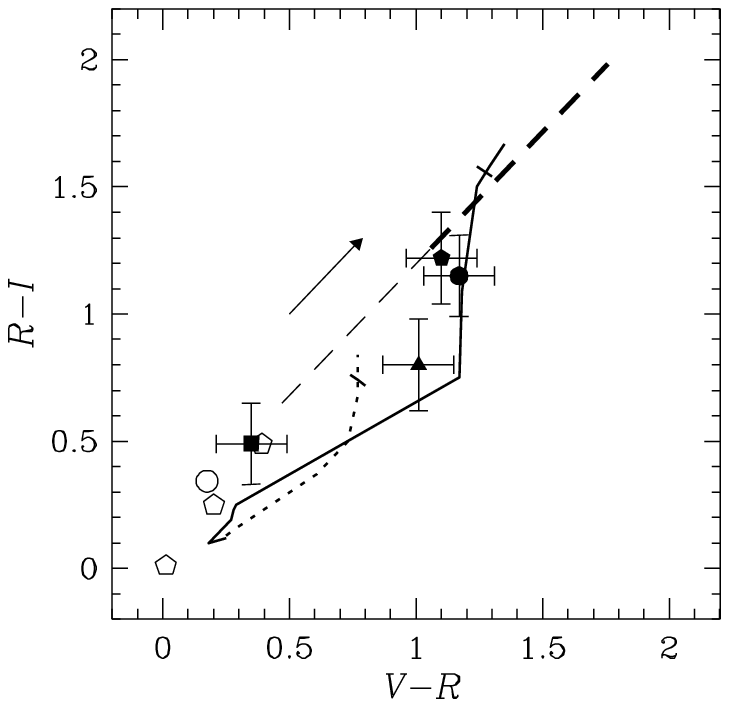} \\
 \caption{}
 \end{figure}
 
 \begin{figure}
 \hspace{2cm}
 \epsfbox{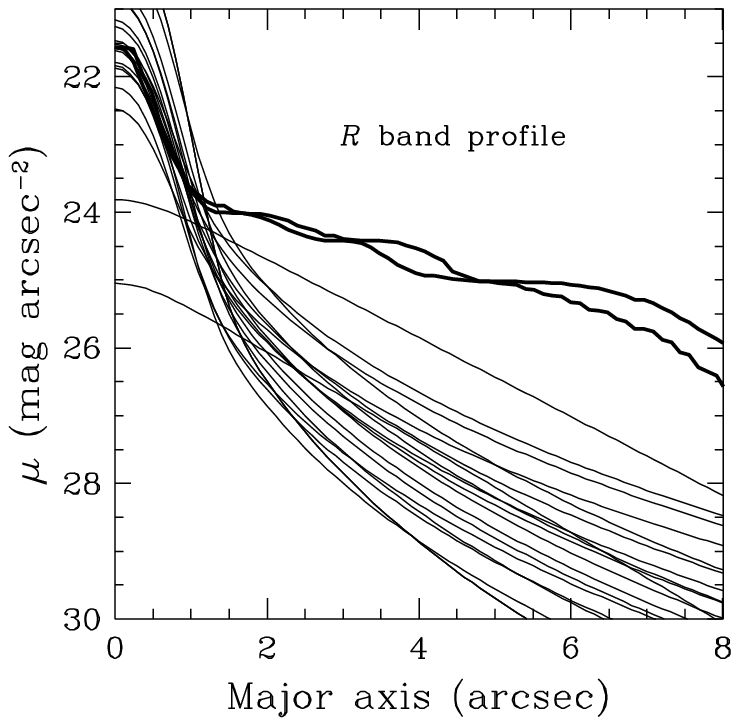} \\
 \caption{}
 \end{figure}
 
 \clearpage

\end{document}